# The Dynamics of Protest Recruitment through an Online Network

Sandra Gonzalez-Bailon[*], Javier Borge-Holthoefer[1], Alejandro Rivero[1], Yamir Moreno[1,2]

[*] Oxford Internet Institute

University of Oxford

1 St. Giles

OX1 3JS, Oxford, UK

[1] Institute for Biocomputation and Physics of Complex Systems

University of Zaragoza

Campus Rio Ebro

50018, Zaragoza, Spain

[2] Department of Theoretical Physics, Faculty of Sciences,

University of Zaragoza, Zaragoza 50009, Spain.

Corresponding author: Dr. Sandra Gonzalez-Bailon, Oxford Internet Institute, University of Oxford, 1 St. Giles, Oxford, OX1 3JS, telephone: +44 (0) 1865 287 233, fax: +44 (0) 1865 287 211, e-mail: sandra.gonzalezbailon@oii.ox.ac.uk.

**The recent wave of mobilizations in the Arab world and across Western countries has generated much discussion on how digital media is connected to the diffusion of protests. We examine that connection using data from the surge of mobilizations that took place in Spain in May 2011. We study recruitment patterns in the Twitter network and find evidence of social influence and complex contagion. We identify the network position of early participants (i.e. the leaders of the recruitment process) and of the users who acted as seeds of message cascades (i.e. the spreaders of information). We find that early participants cannot be characterized by a typical topological position but spreaders tend to me more central to the network. These findings shed light on the connection between online networks, social contagion, and collective dynamics, and offer an empirical test to the recruitment mechanisms theorized in formal models of collective action.**

The last few years have seen an eruption of political protests aided by internet technologies. The phrase "Twitter revolution" was coined in 2009 to refer to the mass mobilizations that took place in Moldova and, a few months later, in Iran, in both cases to protest against fraudulent elections. Since then, the number of events connecting social media with social unrest has multiplied, not only in the context of authoritarian regimes – exemplified by the recent wave of upsurges across the Arab world – but also in western liberal democracies, particularly in the aftermath of the financial crisis and changes to welfare policies. These protests respond to very different socio-economic circumstances and are driven by very different political agendas, but they all seem to share the same morphological feature: the use of social networking sites (SNSs) to help protesters self-organize and attain a critical mass of participants. There is, however, not much evidence on how exactly SNSs encourage recruitment. Empirical research on online activity around riots and protests is scarce, and the few studies that exist [1-4] show no clear patterns of protest growth. Related research has shown that information cascades in online networks occur only rarely [5-7], with the implication that even online it is difficult to reach and mobilize a high number of people. Revolutions, riots and mass mobilizations are also statistically unique and, as such, difficult to predict; but when they happen, they unleash potentially dramatic consequences. The relevant question, which we set to answer here, is not when these protests take place but whether and how SNSs contribute to trigger their explosion.

Sociologists have long analyzed networks as the main recruitment channels through which social movements grow [8-9]. Empirical research has shown that networks were crucial to the organization of collective action long before the internet could act as an organizing tool, with historical examples that include the insurgency in the Paris commune of 1871 [10], the 60s civil right struggles in the U.S. [11], and the demonstrations that took place in East Germany prior to the fall of the Berlin wall [12-13]. These studies provide evidence that recruits to a movement tend to be connected to others already involved and that networks open channels through which influence on behavior spreads, but they are limited by the quality of the network data analyzed, particularly around time dynamics. Analytical models have tried to overcome these data limitations by recreating the formal features of interpersonal influence, and analyzing how they are related to diffusion [14-17] and, more particularly, to examples of social contagion like collective action or the growth of social movements [18-22]. Four main findings arise from these models. First, the shape of the threshold distribution, i.e. the variance in the propensity to join intrinsic to people, determines the global reach of cascades. Second, individual thresholds interact with the size of local networks: two actors with the same propensity might be recruited at different times if one is connected to a larger number of people. Third, attaining a critical mass depends on being able to activate a sufficiently large number of low threshold actors that are also well connected in the overall network structure. And fourth, the exposure to multiple sources can be more important than multiple exposures: unlike epidemics, the social contagion of behavior often requires reinforcement from multiple people. Recent experiments have confirmed the relevance of complex contagions to explain behavior in online contexts [23], and large-scale analyses have validated its effects on information diffusion on Twitter [24].

Models of collective action have identified important network mechanisms behind the decision to join a protest, but they suffer from lack of empirical calibration and external validity. Online networks, and the role that SNSs play in articulating the growth of protests, offer a great opportunity to explore recruitment mechanisms in an empirical setting. We analyze one such setting by studying the protests that took place in Spain in May 2011. The mobilization emerged as a reaction to the political response to the financial crisis and it organized around broad demands for new forms of democratic representation. The main target of the campaign was to organize a protest on May 15, which brought tens of thousands of people to the streets of 59 cities all over the country. After the march, hundreds of participants decided to camp in the city squares until May 22, the date for local and regional elections; crowded demonstrations took place daily during that week. After the elections, the movement remained active but the protests gradually lost strength and its media visibility waned (more background information in *SI*)

We analyze Twitter activity around those protests for the period April 25 (20 days before the first mass mobilizations) to May 25 (10 days after the first mass mobilizations, and 3 days after the elections). The dataset follows the posting behavior of 87,569 users and tracks a total of 581,750 protest messages (see *Methods*). We know, for each user, who they follow and who is following them. In addition to this asymmetric network, we also consider the symmetric version, which only retains reciprocated – and therefore stronger – connections. Previous research has suggested that Twitter is closer to a news media platform than to a social network [6]. The symmetric network eliminates the relatively higher influence of hubs and retains only connections that reflect offline relationships or, at least, mutual acknowledgement. Contrasting recruitment patterns in both the asymmetric and symmetric networks allows us to test whether the dynamics of mobilization depend on weak, broadcasting links or on stronger connections, based on mutual recognition. Our analysis of recruitment is based on the assumption that users joined the movement the moment they started sending Tweets about it. We also assume that once they are activated, they remain so for the rest of the period we consider.

**Results**

By the end of our 30-day window, most users in the network had sent at least one message related to the protest, with only about 2% remaining silent (but still being exposed to movement information, Fig. 1).The most significant increase in activity takes place right after the initial protest (May 15), during the week leading to the elections of May 22. Up to that point, only about 10% of the users had sent at least a message related to the protests.

Activation times tell us the exact moment when users start emitting messages, and allow us to distinguish between activists leading the protests and those who reacted in later stages. We calculated, for each user, the proportion of neighbors being followed that had been active at the time of recruitment ($k_a / k_{in}$). This gives us a measure that approximates the threshold parameter used in formal models of social contagion, particularly those that incorporate networks [16-17,21]. Activists with an intrinsic willingness to participate have a threshold $k_a / k_{in} \approx 0$, whereas those who need a lot of pressure from their local networks before they decide to join are in the opposite extreme $k_a / k_{in} \approx 1$. Looking at the empirical distribution, most users in our case exhibit intermediate values (Fig.2A). Although the distribution is roughly uniform for almost the full threshold interval, there are two local maxima at 0 (users who act as the recruitment seeds) and 0.5 (the conditional co-operators who join when half their neighbors did). The symmetric network has a significantly higher number of users with $k_a / k_{in} = 0$ because it eliminates the influence of hubs or broadcasters (i.e. users who do not reciprocate connections – about 7,000 in the overall network – but who contribute to activate low threshold participants, the 'seeds' in the symmetric network). The shape of the distribution changes before and after 15 May, the first big demonstration day (Fig. 2B). Most early participants – i.e. users who sent a message prior to the first mass mobilizations and to the news media coverage of the events – needed, on average, less local pressure to join, which is consistent with their role as leaders of the movement. Because most activity takes place after 15-M, the threshold distribution for the ten days that followed is not very different from the threshold distribution for the full period.

The actual chronological time of activation changes across same-threshold actors (see *SI,* Fig. S2); this variance is predictable given that actors react to different local networks, both in size and composition, creating an interaction effect that has already been studied using simulations [16-17,21-22]. The time it takes neighbors to join, however, also influences the activation of users. We measure the pace at which the number of active neighbors grows using the logarithmic derivative of activation times $\Delta k_a/k_a = (k_a^{t+1} - k_a^t) / k_a^{t+1}$ [25]. The rationale behind $\Delta k_a/k_a$ is that some users might be susceptible to "recruitment bursts", that is, more likely to join if many of their neighbors do in a short time-span. This emphasis on time dynamics qualifies the idea of complex contagion: receiving stimuli from multiple sources is important because, unlike epidemics, social contagion often requires exposure to a diversity of sources [21]; evidence of "recruitment bursts" would suggest that the effects of multiple and diverse exposures are magnified if they take place in a short time window. We find that early participants, i.e. users with low thresholds, are insensitive to recruitment bursts; for the vast majority of users, however, being exposed to sudden rates of activation precedes their decision to join (Fig. 3A). Users with moderate thresholds who are susceptible to bursts act as the critical mass that makes the movement grow from a minority of early participants to the vast majority of users: without them, late participants (the majority of users that made the movement explode) would not have joined in (Fig. 3B).

Information diffusion follows different dynamics. Very few messages generate cascades of a global scale: we assume that if a user emits a message at time $t$ and one of their followers also emits a message within the interval ($t$, $t+ \Delta t$), both messages belong to the same chain. A chain is aborted when none of the followers exposed to a message acts as a spreader, and messages can only belong to a single chain, i.e. only the messages that do not belong to a previous chain are considered seeds for a new cascade (see *Methods*). The vast majority of these chains die soon, with only a very small fraction reaching global dimensions, a result that is robust using different time intervals (Fig. 4A). This supports previous findings [5-7] and reveals that cascades are rare even in the context of exceptional events. We run a *k*-shell decomposition analysis [26] to identify the network position of users acting as seeds of the most successful chains. We found a positive association between network centrality, as measured by the classification of nodes in high *k*-cores, and cascade size (Fig. 4B). This positive association suggests that agents at the core of the network – not necessarily those with a higher number of connections, but connected to equally well connected users (Fig. 4C-D) – are the most effective when it comes to spreading information, again in accordance with what has been found in research on epidemics and contagion [27]. Spreaders, though, need to be recruited first, and the same decomposition analysis does not find any significant association between thresholds and topology, i.e. early participants do not have a characteristic network position; they are instead scattered all over the network (see *SI*, Fig. S5).

**Discussion**

The role that SNSs play in helping protests grow is uncontested by most media reports of recent events. However, there is not much evidence of how exactly these online platforms can help disseminate calls for action and organize a collective movement. Our findings suggest that there are two parallel processes taking place: the dynamics of recruitment, and the dynamics of information diffusion. While being central in the network is crucial to be influential in the diffusion process, there is no topological position that characterizes the early participants that trigger recruitment. This suggests that whatever exogenous factors motivate early participants to start sending messages, the consequence is that they create random seeding in the online network: they spur focuses of early activity that are topologically heterogeneous and that spread through low threshold individuals. This finding is consistent with previous work using simulations that test (and challenge) the influential hypothesis [16-17]. However, a small core of central users is still critical to trigger chains of messages of high orders of magnitude. The advantage that this minority has as cascade generators – or as the 'percolating cluster' (16) – derives from their location at the core of the network; this suggests that, contrary to what has been argued in previous research (4), centrality in the network of followers is still a meaningful measure of influence in online networks – at least in the context of mass mobilizations.

The decision to join a protest depends on multiple reasons that we do not capture with online data – for instance, the amount of offline news media to which users are exposed. It is not surprising, then, that network position does not account for time of activation as it does for cascading influence (the diffusion of messages is, for the most part, endogenous and dependant on the network structure). However, there is one element in the recruitment process that is endogenous as well, and that is the timing of exposures. The existence of recruitment bursts indicates that the effects of complex contagion [16] are boosted by accelerated exposure, that is, by multiple stimuli received from different sources that take place within a small time window. These bursts – facilitated by the speed at which information flows online – provide empirical evidence of what scholars of social movements have called, metaphorically, "collective effervescence". We provide a measure for that metaphor and find that most users are susceptible to it. These findings qualify threshold models of collective action that do not take into account the urgency to join that bursts of activity instill in people.

In addition, this study provides evidence of why horizontal organizations (like the platform coordinating this protest, see *SI*) are so successful at mobilizing people through SNSs: their decentralized structure, based on coalitions of smaller organizations, plant activation seeds randomly at the start of the recruitment process, which maximizes the chances of reaching a percolating core; users at this core, in turn, contribute to the growth of the movement by generating cascades of messages that trigger new activations, and so forth. These joint dynamics illustrate the trade-off between global bridges (controlled by well connected users) and local networks: the former are efficient at transmitting information, the later at transmitting behaviors (21). This is one reason why Twitter is behind so many recent protests and mobilizations: it combines the global reach of broadcasters with local, personalized relations (which we capture in the form of reciprocal connections); in the light of our data, both are important to articulate the growth of a movement. Again, being able to generate recruitment patterns on a scale of this order is still an exceptional event, and this study sheds no light that helps predict when it will happen again; but it shows that when exceptional events like mass mobilizations take place, recruitment and information diffusion dynamics are reinforcing each other along the way.

Our data has two main limitations. First, we might be overestimating social influence because we do not control for demographic information and the effects of homophily in network formation [28]. Studies that control for demographic attributes, however, still find that networks are significant predictors of recruitment [9,13]; in the light of those findings, we can only assume that online networks will still be significant channels for the spread of behavior once demographics are taken into account. Second, we also do not control for exposure to offline media, which is likely to have interacted with social influence. However, the lack of media coverage before the demonstrations of May 15 allows us to conduct a natural experiment and compare how the network channels recruitment with and without the common knowledge of media exposure. We show that there is no significant shift to the left of the threshold distribution once the media starts reporting on the protests; this would have indicated that exposure to mass media led to a higher proportion of users joining the protests in the absence of local pressure. On the contrary, we find that local pressure is still an important precursor for a large number of users, and that the vast majority are still susceptible to bursts of activity in their local networks.

Future research should address if the dynamics we identify with Twitter data are platform-dependent or universal to different types of online networks. Recent events, like the riots in London in August 2011, suggest that different online platforms are being used to mobilize different populations. The question that future research should consider is if the same recruitment patterns apply regardless of the technology being used, or if the affordances of the technology (i.e. public/private by default) shape the collective dynamics that they help coordinate. The replication of these analyses with data covering similar events (like the OccupyWallStreet protests initiated in New York, and now spreading to other U.S. cities) will also help determine if the dynamics we identify here can be generalized to different social contexts.

**Methods**

The data contains time-stamped tweets for the period April 25 to May 25. Messages related to the protests were identified using a list of 70 #hashtags (full list in *SI*). The collection of messages is restricted to Spanish language and to users connected from Spain, and it was archived by a local start-up company, *Cierzo Development Ltd* using the SMMART Platform. We estimate that our sample captures above a third of the total number of messages exchanged in Twitter related to the protests. The network of followers was reconstructed applying a one-step snowball sampling procedure, using the authors that sent protest messages as the seed nodes. An arc ($i,j$) in this network means that user $i$ is following the Tweets of user $j$, and we assume that this network is static for the period we consider. The symmetric network filters out all asymmetric arcs, that is, for every arc ($i,j$) there also needs to be an arc ($j,i$).

We reconstruct message chains assuming that protest activity is contagious if it takes place in short time windows. We do not have access to re-tweet (RT) information, but since all our messages are related to the 15-M movement, chains refer to the same subject matter (although the precise content of the messages in the same chain might differ). This measurement maps the extent to which the stream of content related to the protests diffuses in given time windows.

The *k*-shell decomposition assigns a shell index $k_s$ to each user by pruning the network down to users with more than *k* neighbours. The process starts removing all nodes with degree $k = 1$, which are classified (together with their links) in a shell with index $k_s = 1$. Nodes in the next shell, with degree $k = 2$, are then removed and assigned to $k_s = 2$, and so forth until all nodes are removed (and all users are classified). Shells are layers of centrality in the network: users classified in shells with higher indexes are located at the core, whereas users with lower indexes define the periphery of the network (see *SI* for details of node classification in shells).

**Acknowledgements**

J. B-H is partially supported by the Spanish MICINN through project FIS2008-01240. Y. M. is supported by the Spanish MICINN through projects FIS2008-01240 and FIS2009-13364-C02-01 and by the Government of Aragon (DGA) through the grant No. PI038/08.

**Authors Contributions**

S. G-B, J. B-H and Y.M. designed research, analyzed data, and wrote the paper. A.R. analyzed data.

**Additional Information**

Competing Financial Interests

The authors declare no competing financial interests.

**Figure legends**

Fig 1. Fraction of recruited users over time. The vertical axis is normalized by the total number of users (87,569), the horizontal axis tracks the number of activated users accumulated by hours. At the end of our time window the proportion of activated users is 98.03%, which means that the vast majority of users sent at least one protest message during this month. Vertical labels flag some of the events that took place during the period.

Fig 2. Distribution of thresholds $k_a / k_{in}$. (A) The vertical axis measures the proportion of users activated for each threshold of activated neighbors, tracked in the horizontal axis. The figure shows measures for both the asymmetrical and symmetrical networks. When broadcasters (i.e. central users who do not reciprocate connections) are eliminated, the number of early participants with $k_a / k_{in} = 0$ increases by an order of magnitude, which suggests that broadcasters are influential at recruiting low-threshold individuals. Panel (B) splits the data in two subsets: the first subset considers recruitment activity before 15-M, the day of the first mass demonstrations; the second subset tracks activity after 15-M. Media coverage, which increased after 15-M, does not seem to cause a significant rise in the number of early activated, low-threshold users.

Fig 3. Thresholds, recruitment bursts, and time of activation. (A) The figure measures the association between bursts of activity and thresholds; while early participants ($k_a / k_{in} < 0.2$) are not affected by bursts, moderate-threshold users ($0.2 \leq k_a / k_{in} \leq 0.5$) and high threshold users ($k_a / k_{in} > 0.5$) are more likely to join the exchange of messages if they see a sudden increase of participants in their local networks; the slope of the curve indicates that higher threshold users are more susceptible to bursts of activity. (B) This figure shows the percentage of activated users grouped as early, mid and late participants for each day of the period considered; most late participants joined the protests after 15-M, once a critical mass of mid participants had already been activated.

Fig 4. Distribution of cascade sizes and core position of spreaders. (A) The distribution of cascade sizes ($N_c$) suggests that only a few cascades percolate to affect most users, and that the vast majority die in the early stages of diffusion. (B) There is a positive correlation between the *k*-core of users starting the cascades, and the size of those cascades, suggesting that core users are more likely to be the seeds of global chains of information diffusion. (C) The nodes in the network arranged according to their *k*-core: higher *k*-cores indicate that nodes are more central, node size accounts for degree centrality, and node color indicates the maximum size of the cascades generated by the user (users generating the largest cascades are depicted in orange). (D) Example of a global cascade affecting about 35,000 nodes. Nodes in blue are users who participated in the diffusion of protest messages; nodes in orange were exposed to the messages but did not send messages of their own. The darker the shade of blue, the earlier users joined the cascade as spreaders; the lighter the shade of yellow, the later users joined the cascade as listeners.

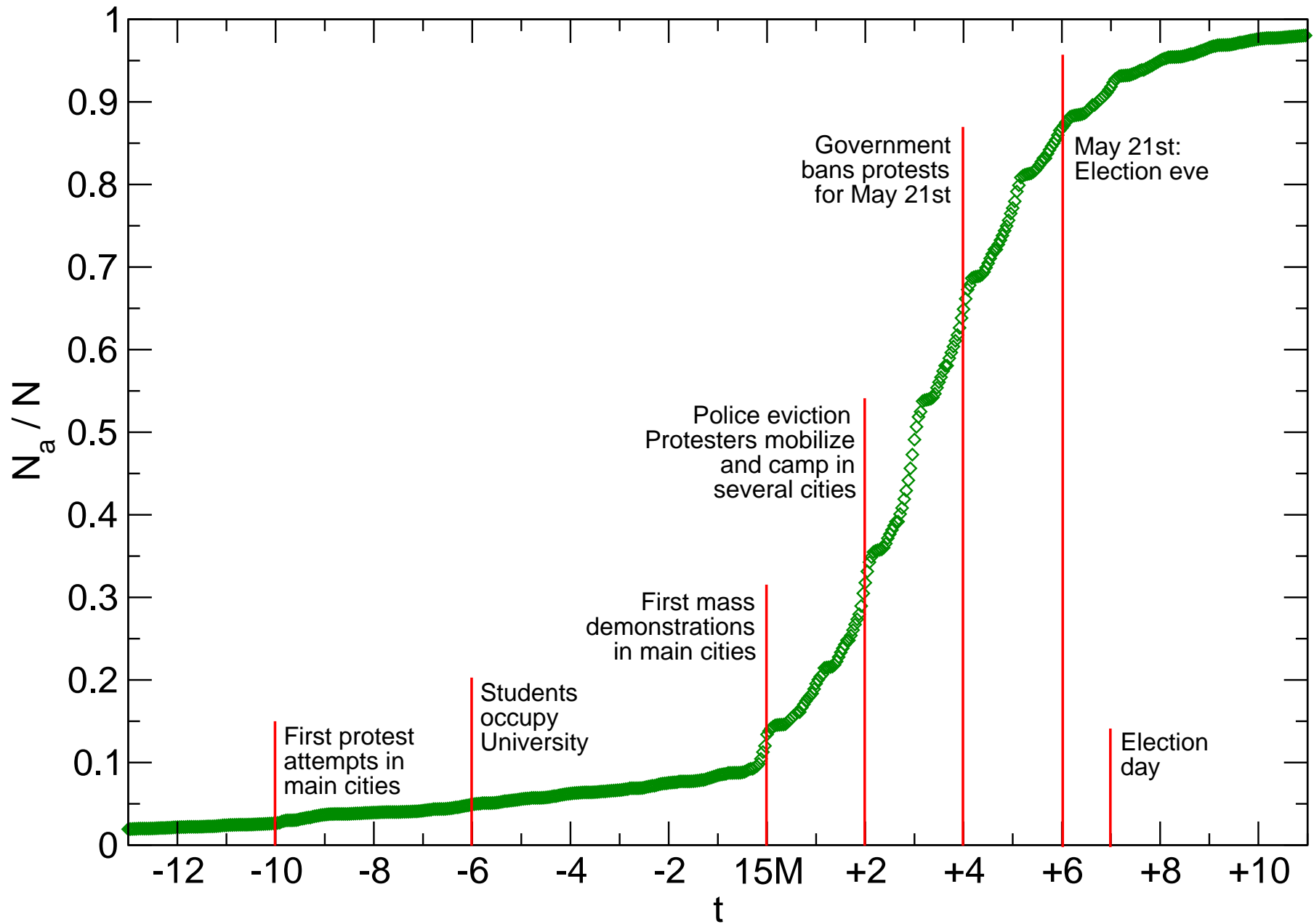

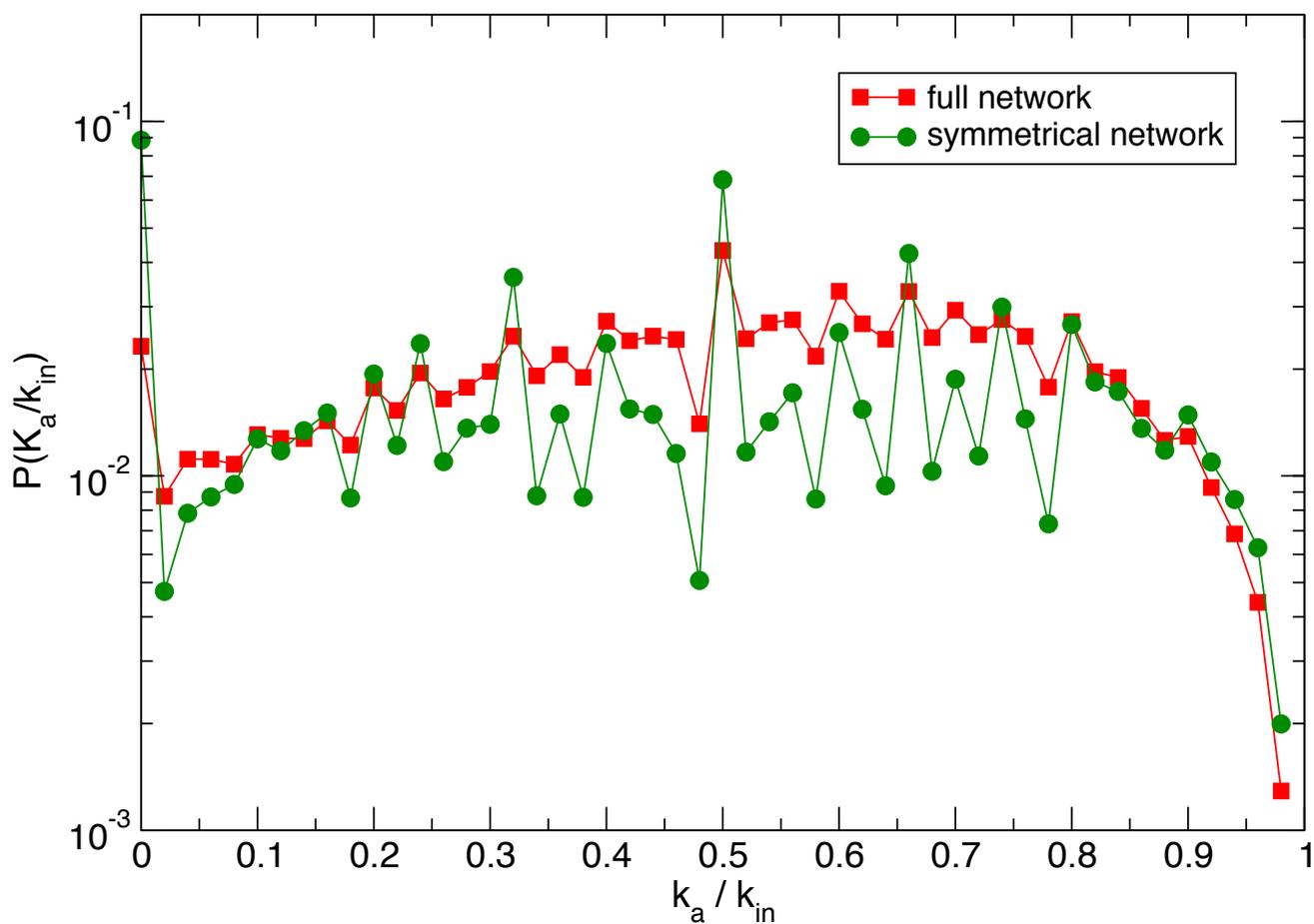

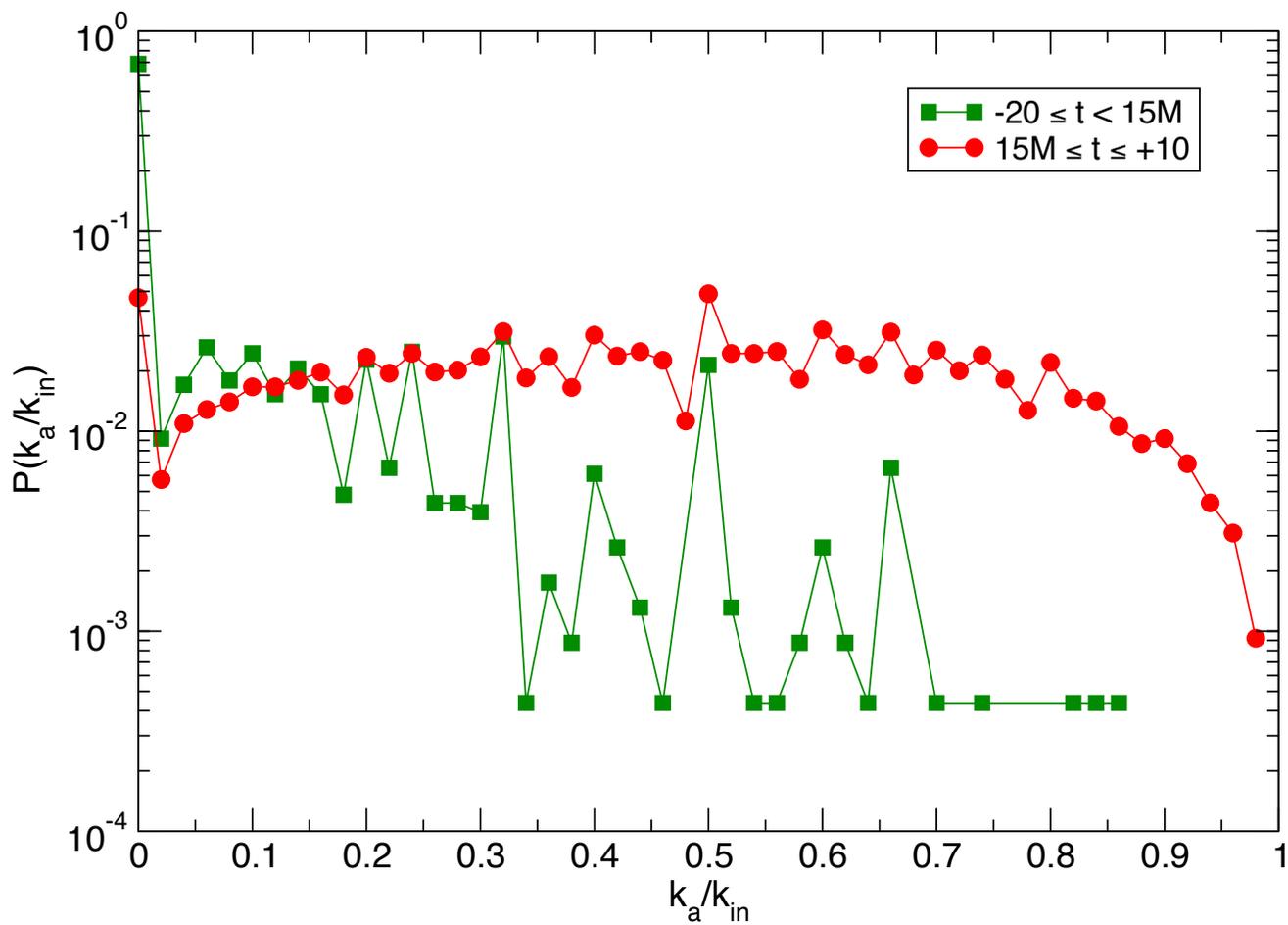

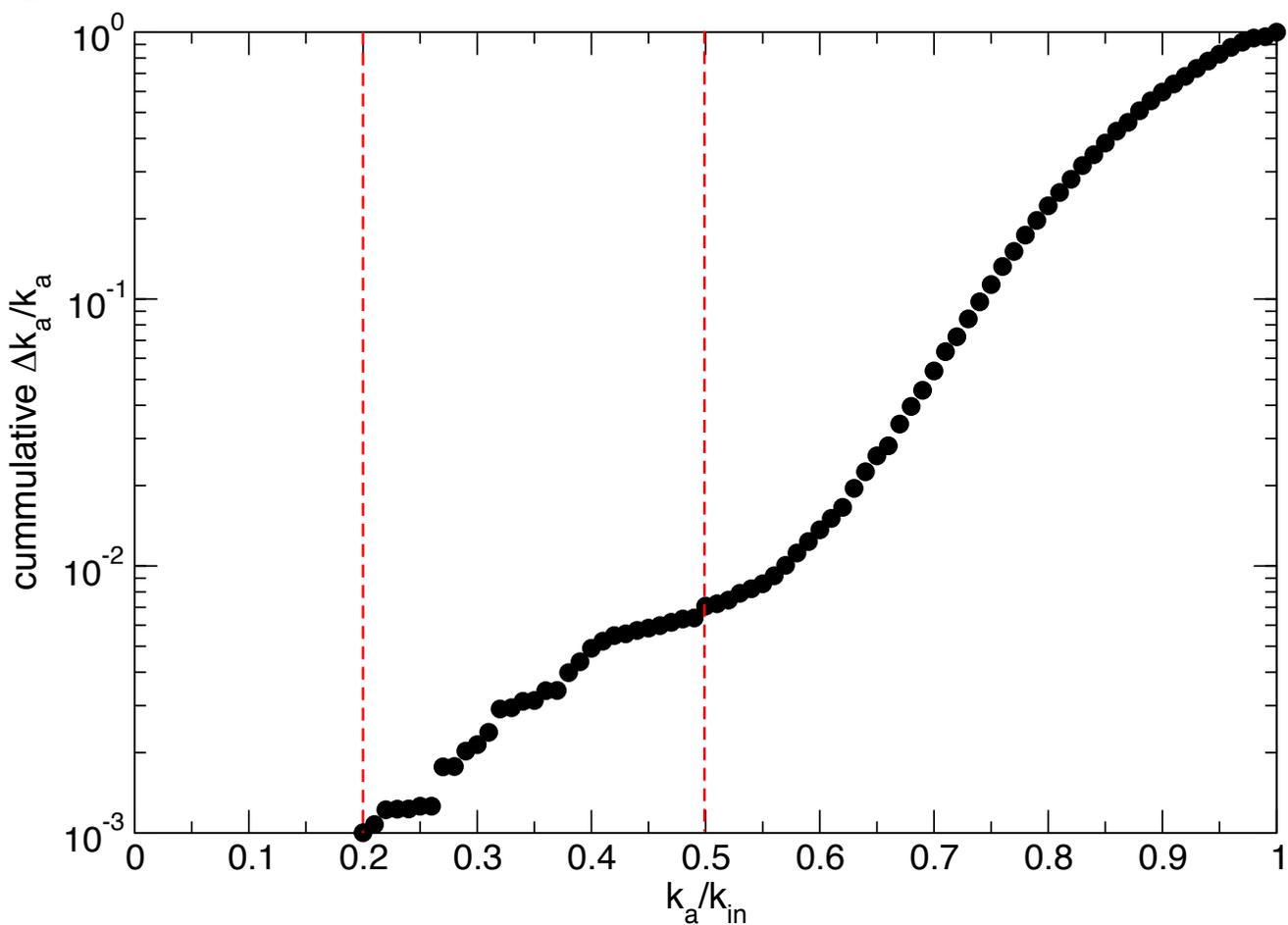

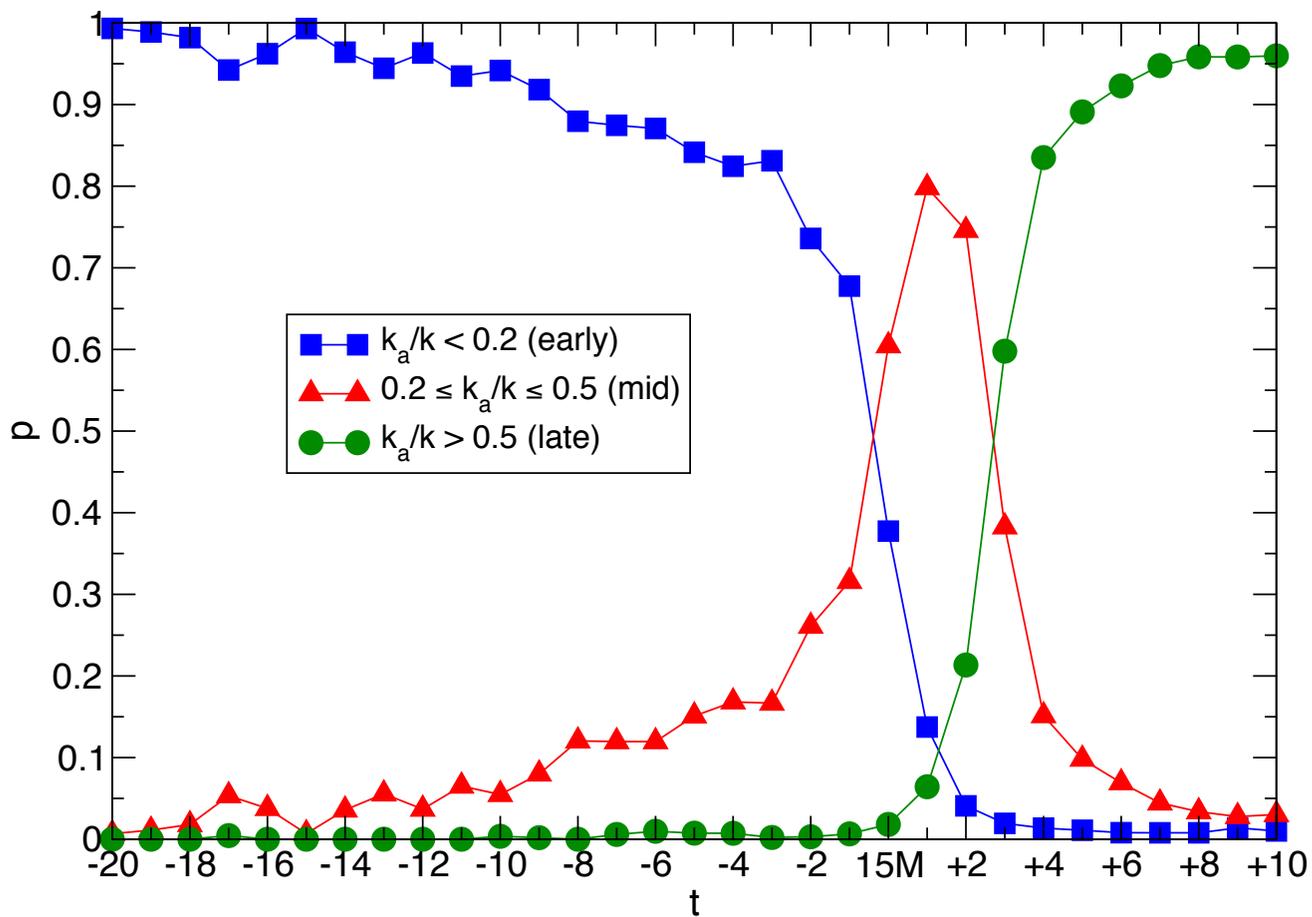

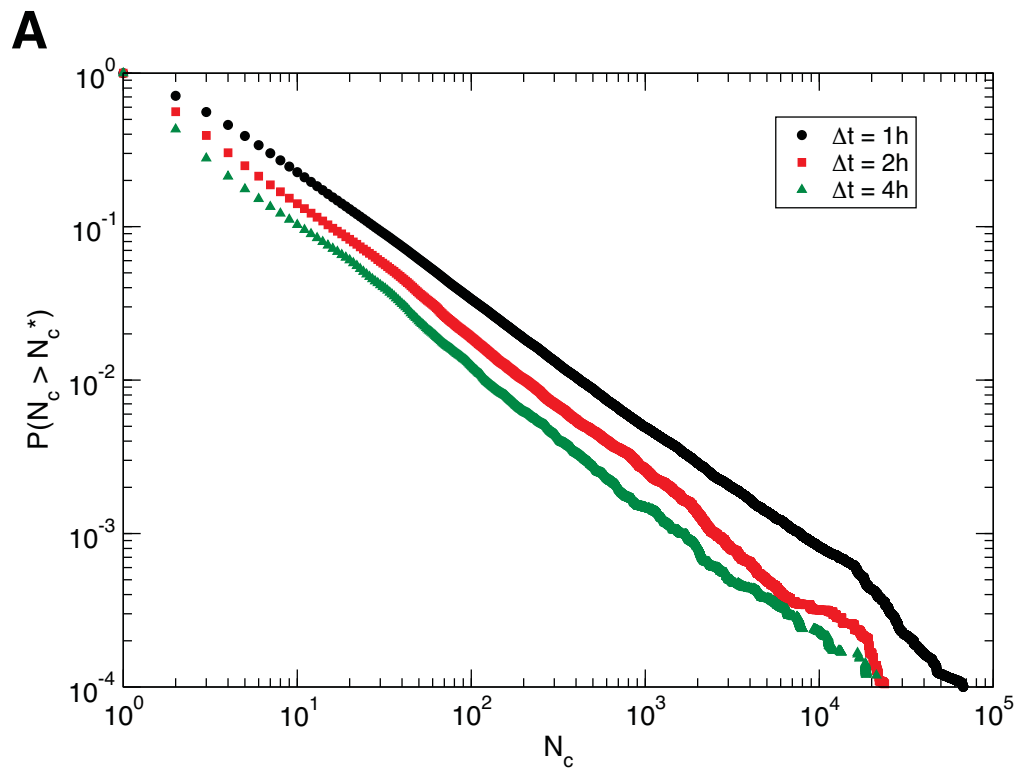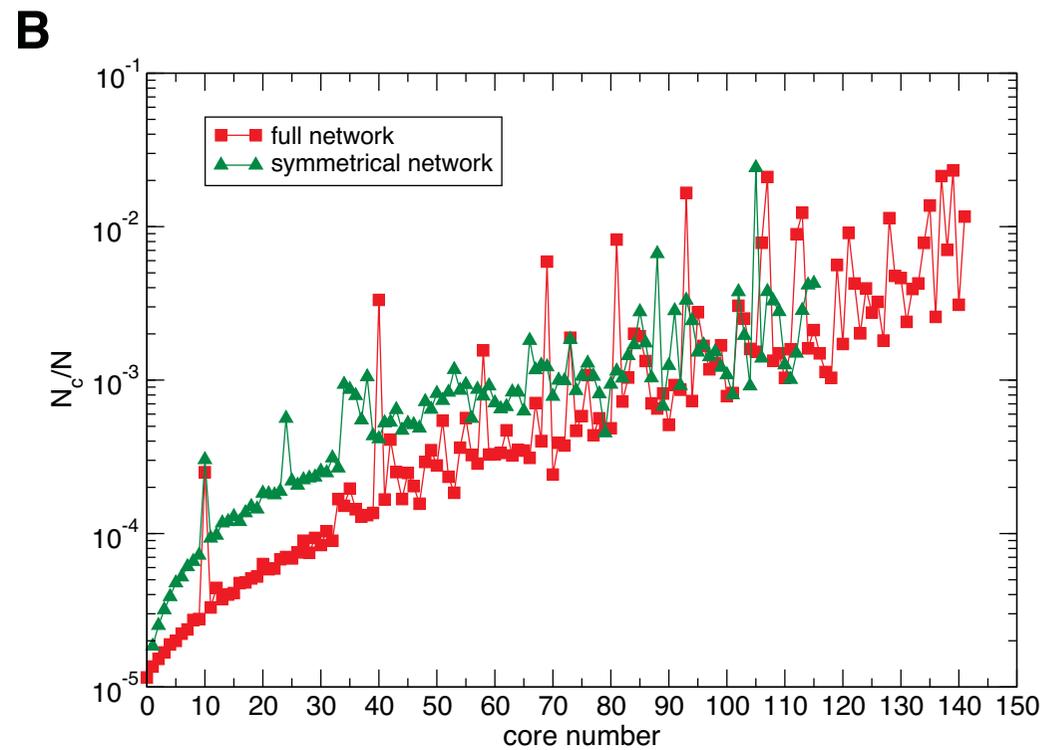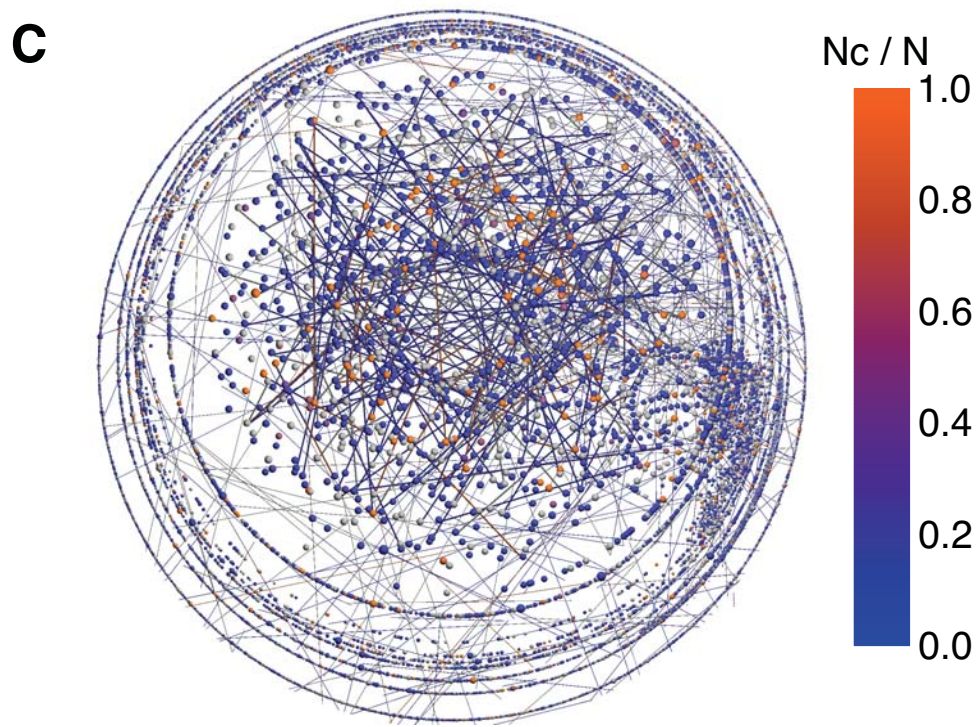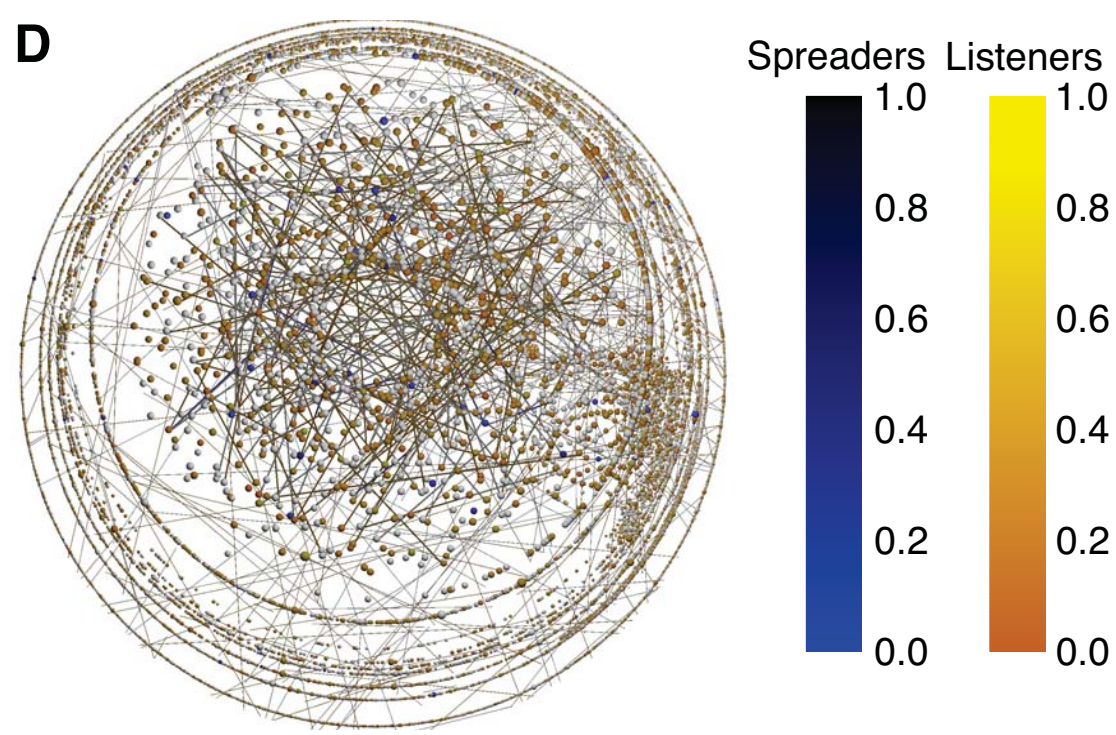

SUPPLEMENTARY INFORMATION

**The Dynamics of Protest Recruitment through an Online Network**

Sandra Gonzalez-Bailon[*], Javier Borge-Holthoefer[†], Alejandro Rivero[†], Yamir Moreno[†‡]

[*] Oxford Internet Institute

University of Oxford

1 St. Giles

OX1 3JS, Oxford, UK

[†] Institute for Biocomputation and Physics of Complex Systems

University of Zaragoza

Campus Rio Ebro

50018, Zaragoza, Spain

[‡] Department of Theoretical Physics, Faculty of Sciences,

University of Zaragoza, Zaragoza 50009, Spain.

**Protests Background**

The 15-M movement is a civic initiative with no party or union affiliation that emerged as a reaction to perceived political alienation and to demand better channels for democratic representation. The first mass demonstration, held on Sunday May 15, was conceived as a protest against bipartidism and the management of the economy in the aftermath of the financial crisis. It was organized by the digitally coordinated platform "Real Democracy Now" ("Democracia Real Ya"), born online about three months before the first day of demonstrations. Hundreds of entities joined the platform, from small local associations to territorial delegations of larger groups like ATTAC (an international anti-globalization organization) or Ecologists in Action (Ecologistas en Acción). Signatories of the original call included student associations, bloggers, defenders of human rights and people from the arts, but also hundreds of individual citizens of different age and ideologies. Under the motto "take the streets" ("toma la calle"), the movement organized peaceful protests that brought tens of thousands of people to the streets of more than fifty cities all over the country, with Madrid, Barcelona, Málaga, Alicante and Valencia leading in numbers. The slogans of the demonstrations included "we are not goods in the hands of politicians and bankers", "we don't pay this crisis" and "no more corruption, let's pass to action". The organizers cited as inspiring examples the Greek protests of 2008 and the revolutions in the Arab world earlier in 2011.

After the 15-M demonstrations, hundreds of participants decided to continue the protests camping in the main squares of several cities (Puerta del Sol in Madrid, Plaça de Catalunya in Barcelona) until May 22, the following Sunday and the date for regional and local elections. During that week, protesters created committees to coordinate the logistics of the camp sites (from cleaning, to cooking, getting blankets or assembling covers to protect campers from the rain) and organized around open popular assemblies. The media, which had not covered the movement until the day of the first big demonstrations, started covering the protests on a daily basis, particularly after the authorities tried to evict protesters from the squares by force, and the Electoral Committee declared the protests illegal. Despite the prohibition, the camps remained in place, receiving increasing popular support and staging daily demonstrations (see Fig. S1 for timeline). Some of the most popular hashtags during this week included #spanishrevolution, #democraciarealya ("real democracy now"), #nonosvamos ("we don't go"), #15M, and #notenemosmiedo ("we are not afraid"). Although the movement did not support any political party or explicitly advocated for strategic voting, the implicit demand was to support smaller parties to the detriment of the two majority parties, the incumbent social-democratic Socialist Party (PSOE) and the opposition centre-right Popular Party (PP). The elections of May 22, however, gave an overwhelming victory to the PP across the country. On that day, protesters in Madrid and Barcelona decided to stay in the squares. After a few altercations with the police (and increasing opposition from local commerce), protesters eventually vacated the squares in early June, four weeks after the protests started. The movement is still active, with the next big demonstration planned for October 15 2011 under the motto "United for Global Change" (see www.democraciarealya.es).

**Data Collection**

A list of 70 hashtags was used to identify Twitter messages related to the 15-M protests (see Table S1). The top seven hashtags appear in 71% of the messages analyzed. Because the collection of messages is restricted to Spanish language, activity expressed in any of the other official languages of Spain (Catalan, Basque, and Galician) is excluded from the sample, as are messages contributed by users not based in Spain. Many messages used more than one hashtag at a time.

The matrix of followings and followers was collected at the end of the period studied. To collect it, we queried the Twitter API to retrieve the list of followings and followers for every unique user in our sample. We used a custom purpose Python 2.7 program and a MySQL database, used as the central point of data collection. An example of the data mining program is available at https://github.com/Ibercivis/TwitterDataMining/. Currently, the Twitter API has a limit of about 60-120 queries per IP address and hour. We spread the Python data mining program over a cloud of 128 different IPs using Eucaliptus as cloud controller and an address range from the university campus.

Table S2 contains some descriptive statistics for the full and symmetrical networks reconstructed using the procedure above. Figure S3 plots the degree distribution for both networks; the difference is expected given that the symmetrical network eliminates very popular users who do not tend to reciprocate the connections established by their followers.

**Activation Times and Network Structure**

Figure S2 shows that same-threshold actors vary significantly in their chronological activation day. This variance results from differences in local networks (which expose actors to different information contexts) but also from factors that we do not measure in our data, like frequency of use of Twitter. Variance is particularly high for low and high threshold actors, suggesting that exogenous factors are more likely to affect them. Overall, however, low threshold actors are more likely to be early participants than high threshold actors.

Table S3 shows the number of users allocated to each $k$-shell for both networks in descending order. High $k$-values define the core of the network, small values define the periphery. An interesting feature of this distribution is the high number of users classified in the highest $k$-core, compared to the cores preceding it; this indicates that the network has a relatively well-sized core of well-connected users. Figure S4 provides a graphical representation of the $k$-shell decomposition. Nodes are arranged and colored in line with their core position: the further from the centre nodes are, the lower their $k$- core (and the lighter their color); nodes within the same core are arranged according to their degree (captured by node size). The figure shows that the most central users according to the number of connections they have is not always related to their core centrality.

While the vast majority of users are classified in lower *k*-shells, the minority of users that occupy the core of the network are more likely to initiate large cascades; as Figure S5 shows, however, there is no significant association between topological position and being a seed in the recruitment process: users that lead the recruitment process are scattered all over the network.

| Date | Event |
|---|---|
| 25-A | *Start of observation window* |
| 5-M | First demonstration attempts. |
| 9-M | Students occupy university. |
| **15-M** | **First mass demonstrations.** |
| 16-M | Camps in squares of Madrid and Barcelona. |
| 17-M | Police evictions. Protesters mobilise (and camp) in several cities. |
| 18-M | Protests declared illegal. Thousands concentrate in squares. |
| 19-M | Popular support and media coverage increase. |
| 20-M | Thousands camped. |
| 21-M | Tens of thousands protest, defying ban. |
| **22-M** | **Election day.** Protesters decide to stay camped in squares. |
| 25-M | *End of observation window* |

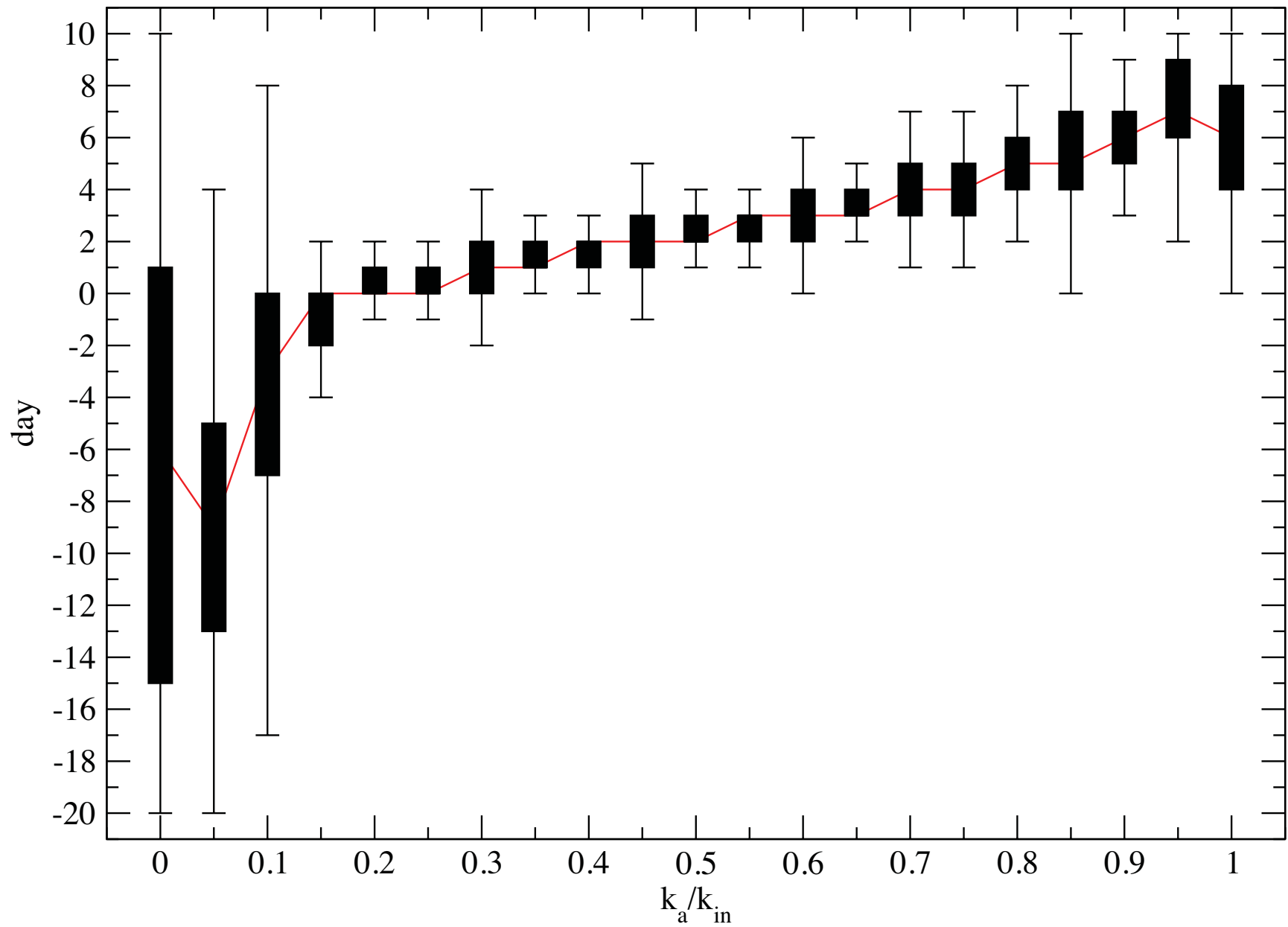

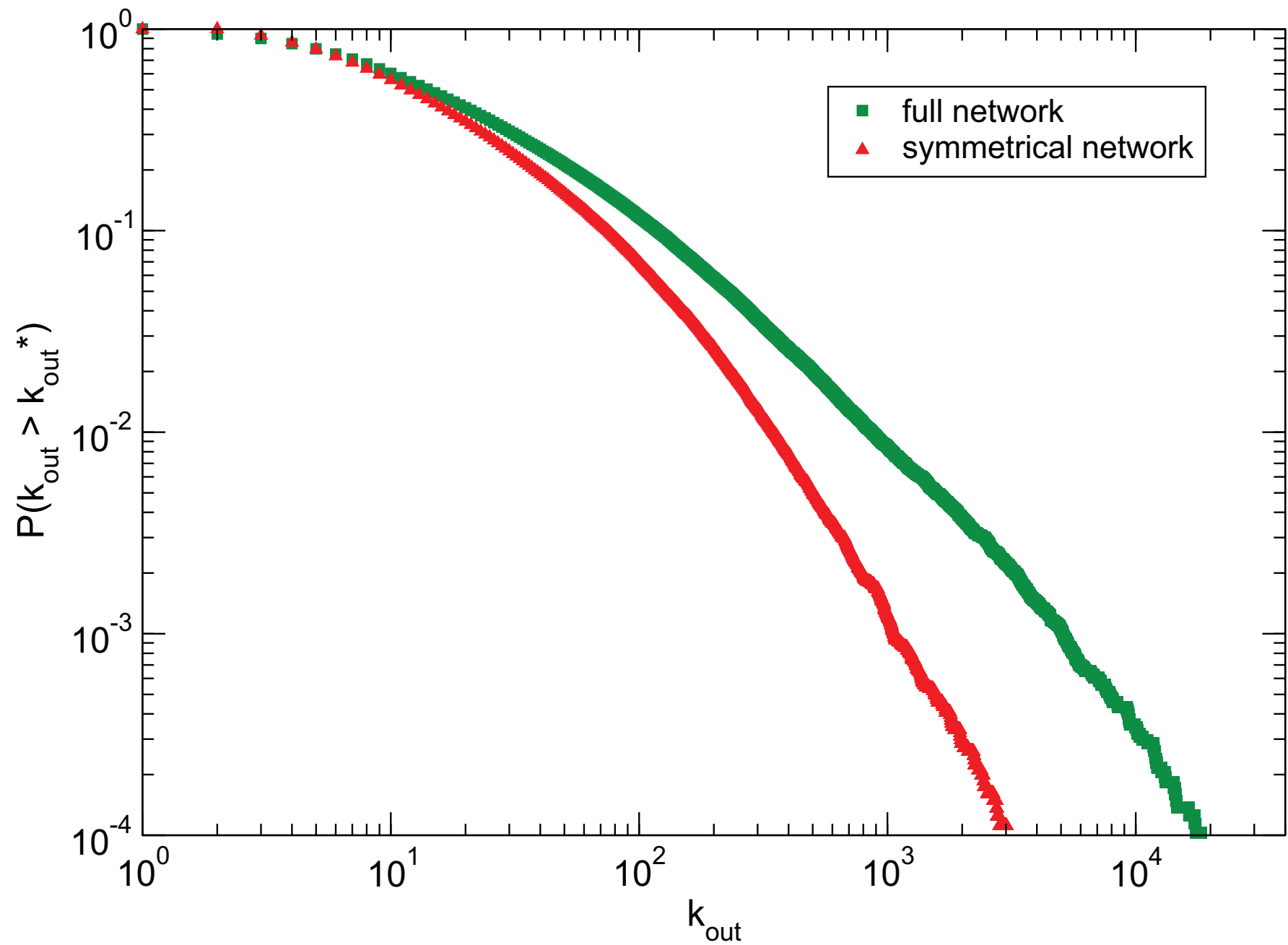

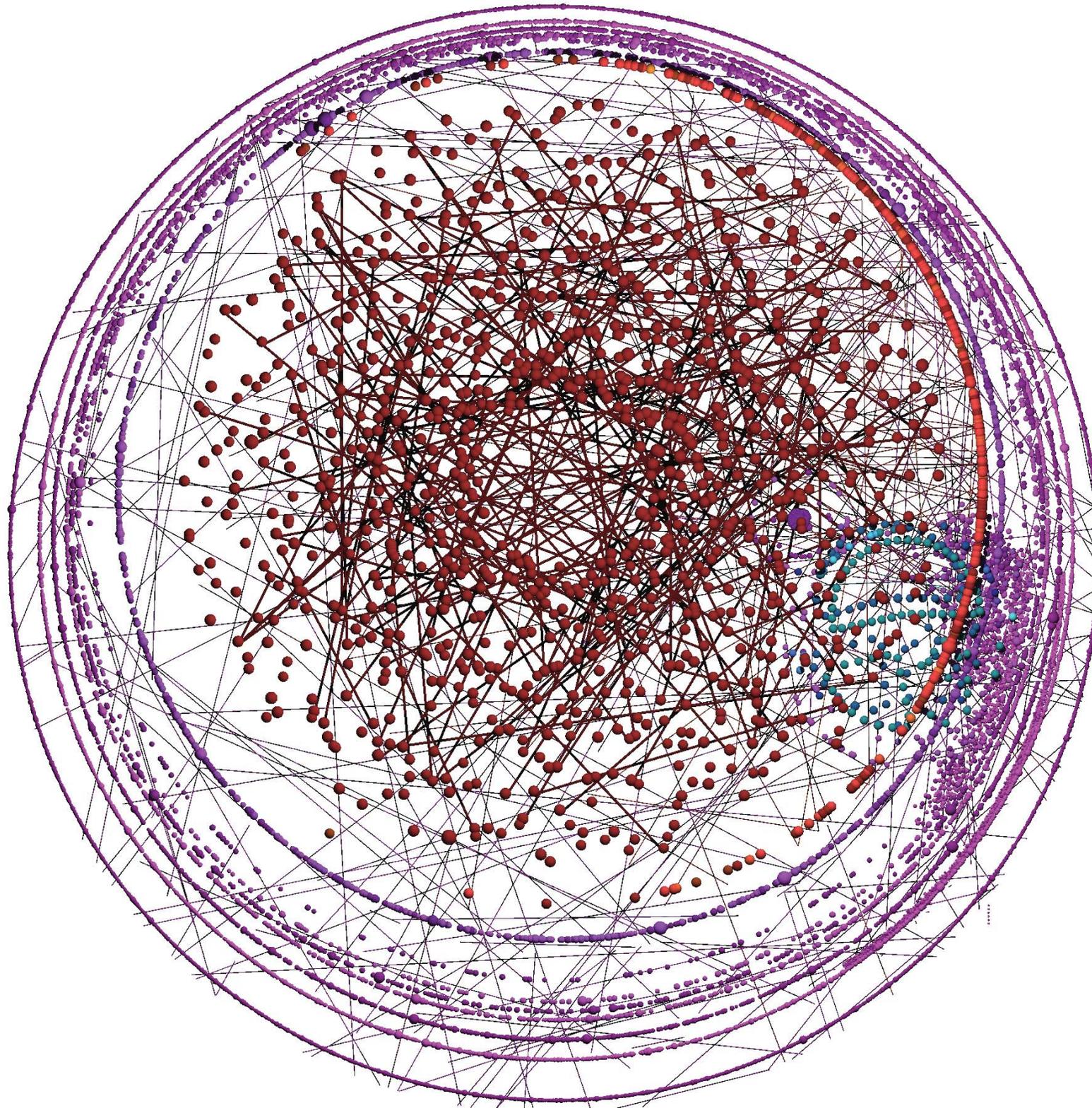

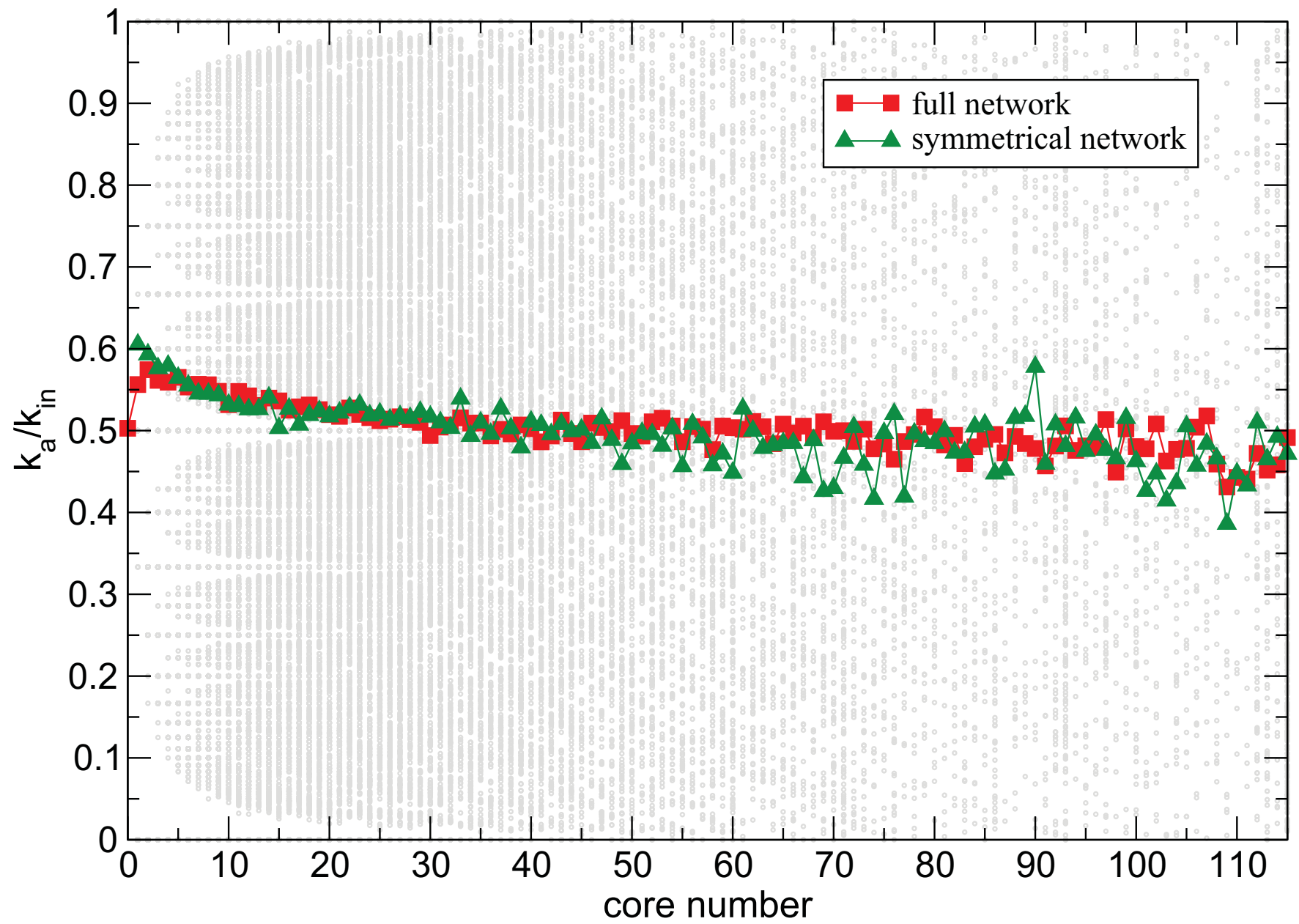

Table S1. Complete list of hashtags used to retrieve Twitter messages related to the 15-M protest.

| | hashtag | translation | number of messages |
|---|---|---|---|
| 1 | #acampadasol | camping Sol | 189251 |
| 2 | #spanishrevolution | Spanish revolution | 158487 |
| 3 | #nolesvotes | don't vote for them | 66329 |
| 4 | #15m | 15-M | 65962 |
| 5 | #nonosvamos | we don't go | 55245 |
| 6 | #democraciarealya | real democracy now | 47463 |
| 7 | #notenemosmiedo | we are not afraid | 32586 |
| 8 | #yeswecamp | yes we camp | 31811 |
| 9 | #15mani | 15 demo | 17986 |
| 10 | #acampadasevilla | camping Seville | 14356 |
| 11 | #globalcamp | global camp | 13186 |
| 12 | #acampadavalencia | camping Valencia | 13129 |
| 13 | #acampadagranada | camping Granada | 9717 |
| 14 | #acampadamalaga | camping Malaga | 6808 |
| 15 | #acampadazgz | camping Zaragoza | 6033 |
| 16 | #consensodeminimos | minimal consensus | 4348 |
| 17 | #italianrevolution | Italian revolution | 3981 |
| 18 | #estonosepara | this doesn't separate | 3860 |
| 19 | #acampadaalicante | camping Alicante | 3593 |
| 20 | #tomalacalle | take the streets | 3517 |
| 21 | #europeanrevolution | European revolution | 3035 |
| 22 | #acampadapamplona | camping Pamplona | 2839 |
| 23 | #worldrevolution | Wold revolution | 2777 |
| 24 | #acampadapalma | camping Palma | 2709 |
| 25 | #tomalaplaza | take the square | 2684 |
| 26 | #acampadas | camps | 2339 |
| 27 | #15mpasalo | 15-M send it | 2336 |
| 28 | #cabemostodas | we all fit in | 1895 |
| 29 | #nonosmovemos | we don't move | 1382 |
| 30 | #3puntosbasicos | 3 basic points | 1378 |
| 31 | #frenchrevolution | French revolution | 1164 |
| 32 | #estonoseacaba | this doesn't end | 1120 |
| 33 | #acampadatoledo | camping Toledo | 750 |
| 34 | #nonosrepresentan | they don't represent us | 696 |
| 35 | #acampadalondres | camping London | 627 |
| 36 | #globalrevolution | global revolution | 622 |
| 37 | #acampadazaragoza | camping Zaragoza | 462 |
| 38 | #acampadaparis | camping Paris | 438 |
| 39 | #takethesquare | take the square | 229 |

Table S1. Complete list of hashtags (continued)

| | hashtag | translation | number of messages |
|---|---|---|---|
| 40 | #periodismoeticoya | ethical journalism now | 207 |
| 41 | #hastalasgenerales | until the general elections | 45 |
| 42 | #irishrevolution | Irish revolution | 39 |
| 43 | #democraziarealeora | real democracy now | 38 |
| 44 | #democraciaparticipativa | participatory democracy now | 33 |
| 45 | #15mpamplona | 15-M Pamplona | 22 |
| 46 | #barcelonarealya | Barcelona real now | 17 |
| 47 | #dry_jaen | real democracy now Jaen | 12 |
| 48 | #usarevolution | USA revolution | 12 |
| 49 | #dry_caceres | real democracy now Caceres | 10 |
| 50 | #dryasturies | real democracy now Asturies | 5 |
| 51 | #democraziareale | real democracy | 4 |
| 52 | #democratiereelle | real democracy | 3 |
| 53 | #dry_cadiz | real democracy now Cadiz | 3 |
| 54 | #dry_toledo | real democracy now Toledo | 3 |
| 55 | #acampadasvlla | camping Seville | 2 |
| 56 | #drybizkaia | real democracy now Vizcaya | 2 |
| 57 | #dry_santander | real democracy now Santander | 2 |
| 58 | #15mayovalencia | 15-M Valencia | 1 |
| 59 | #dry_pisa | real democracy now Pisa | 1 |
| 60 | #dryginebra | real democracy now Geneva | 0 |
| 61 | #DRY_Algeciras | real democracy now Algeciras | 0 |
| 62 | #demorealyaib | real democracy now | 0 |
| 63 | #DRYGipuzkoa | real democracy now Gipuzcua | 0 |
| 64 | #DryValladolid | real democracy now Valladolid | 0 |
| 65 | #ItalRevolution | Italian revolution | 0 |
| 66 | #BolognaDRY | real democracy now Bologna | 0 |
| 67 | #DRY_Pavia | real democracy now Pavia | 0 |
| 68 | #DRY_Almeria | real democracy now Almeria | 0 |
| 69 | #15mayoCordoba | 15-M Cordoba | 0 |
| 70 | #ciudades-dry | real democracy now cities | 0 |

Table S2. Network statistics for the full and symmetrical network

|  | Full Network | Symmetrical Network |
|---|---|---|
| $N$ (# nodes) | 87,569 | 80,715 |
| $M$ (# arcs) | 6,030,459 | 2,644,367 |
| $<k>$ (avg degree) | 69 | 33 |
| $C$ (clustering) | 0.220 | 0.198 |
| $l$ (path length) | 3.24 | 3.65 |
| $D$ (diameter) | 11 | 11 |
| $r$ (assortativity) | -0.139 | -0.0344 |
| # strong components | 5,249 | 139 |
| $N$ giant component | 82,253 | 80,421 |
| N $2^{nd}$ component | 4 | 4 |
| $\max(k_{in})$ (# following) | 5,773 | 5,082 |
| $\max(k_{out})$ (# followers) | 31,798 | 5,082 |

Table S3. Size of k-shells for asymmetrical and symmetrical networks

| asymmetrical network | | symmetrical network | |
|---|---|---|---|
| $k_s$ | N | $k_s$ | N |
| 141 | 1308 | 115 | 471 |
| 140 | 42 | 114 | 55 |
| 139 | 95 | 113 | 88 |
| 138 | 107 | 112 | 106 |
| 137 | 47 | 111 | 14 |
| 136 | 98 | 110 | 16 |
| 135 | 61 | 109 | 24 |
| 134 | 41 | 108 | 35 |
| 133 | 87 | 107 | 33 |
| 132 | 115 | 106 | 45 |
| 131 | 81 | 105 | 52 |
| 130 | 69 | 104 | 44 |
| 129 | 136 | 103 | 15 |
| 128 | 137 | 102 | 39 |
| 127 | 138 | 101 | 37 |
| 126 | 58 | 100 | 78 |
| 125 | 123 | 99 | 48 |
| 124 | 217 | 98 | 31 |
| 123 | 158 | 97 | 91 |
| 122 | 240 | 96 | 53 |
| 121 | 114 | 95 | 61 |
| 120 | 74 | 94 | 35 |
| 119 | 88 | 93 | 174 |
| 118 | 77 | 92 | 97 |
| 117 | 79 | 91 | 50 |
| 116 | 98 | 90 | 46 |
| 115 | 127 | 89 | 43 |
| 114 | 121 | 88 | 65 |
| 113 | 127 | 87 | 16 |
| 112 | 104 | 86 | 96 |
| 111 | 71 | 85 | 55 |
| 110 | 125 | 84 | 66 |
| 109 | 179 | 83 | 73 |
| 108 | 153 | 82 | 48 |
| 107 | 117 | 81 | 48 |
| 106 | 115 | 80 | 68 |
| 105 | 103 | 79 | 55 |
| 104 | 274 | 78 | 68 |
| 103 | 167 | 77 | 58 |

Table S3. Size of k-shells for asymmetrical and symmetrical networks (continued)

| asymmetrical network | | symmetrical network | |
|---|---|---|---|
| $k_s$ | N | $k_s$ | N |
| 102 | 150 | 76 | 106 |
| 101 | 193 | 75 | 80 |
| 100 | 164 | 74 | 54 |
| 99 | 211 | 73 | 81 |
| 98 | 142 | 72 | 88 |
| 97 | 234 | 71 | 182 |
| 96 | 186 | 70 | 103 |
| 95 | 179 | 69 | 98 |
| 94 | 180 | 68 | 86 |
| 93 | 163 | 67 | 94 |
| 92 | 249 | 66 | 143 |
| 91 | 166 | 65 | 104 |
| 90 | 214 | 64 | 117 |
| 89 | 214 | 63 | 107 |
| 88 | 231 | 62 | 139 |
| 87 | 173 | 61 | 119 |
| 86 | 272 | 60 | 189 |
| 85 | 231 | 59 | 140 |
| 84 | 260 | 58 | 216 |
| 83 | 256 | 57 | 153 |
| 82 | 211 | 56 | 152 |
| 81 | 187 | 55 | 164 |
| 80 | 220 | 54 | 160 |
| 79 | 244 | 53 | 180 |
| 78 | 266 | 52 | 263 |
| 77 | 242 | 51 | 161 |
| 76 | 292 | 50 | 222 |
| 75 | 284 | 49 | 264 |
| 74 | 304 | 48 | 221 |
| 73 | 331 | 47 | 230 |
| 72 | 302 | 46 | 210 |
| 71 | 325 | 45 | 235 |
| 70 | 276 | 44 | 320 |
| 69 | 337 | 43 | 341 |
| 68 | 314 | 42 | 297 |
| 67 | 379 | 41 | 355 |
| 66 | 316 | 40 | 321 |
| 65 | 307 | 39 | 391 |
| 64 | 380 | 38 | 372 |
| 63 | 352 | 37 | 371 |

Table S3. Size of k-shells for asymmetrical and symmetrical networks (continued)

| asymmetrical network | | symmetrical network | |
|---|---|---|---|
| $k_s$ | N | $k_s$ | N |
| 62 | 403 | 36 | 414 |
| 61 | 398 | 35 | 433 |
| 60 | 400 | 34 | 457 |
| 59 | 368 | 33 | 426 |
| 58 | 422 | 32 | 485 |
| 57 | 414 | 31 | 461 |
| 56 | 381 | 30 | 526 |
| 55 | 436 | 29 | 609 |
| 54 | 430 | 28 | 524 |
| 53 | 500 | 27 | 659 |
| 52 | 506 | 26 | 748 |
| 51 | 497 | 25 | 831 |
| 50 | 512 | 24 | 738 |
| 49 | 600 | 23 | 850 |
| 48 | 535 | 22 | 850 |
| 47 | 597 | 21 | 916 |
| 46 | 628 | 20 | 956 |
| 45 | 573 | 19 | 1056 |
| 44 | 627 | 18 | 1044 |
| 43 | 597 | 17 | 1267 |
| 42 | 593 | 16 | 1318 |
| 41 | 653 | 15 | 1471 |
| 40 | 693 | 14 | 1628 |
| 39 | 720 | 13 | 1717 |
| 38 | 718 | 12 | 1969 |
| 37 | 822 | 11 | 2300 |
| 36 | 742 | 10 | 2578 |
| 35 | 801 | 9 | 2801 |
| 34 | 799 | 8 | 3193 |
| 33 | 853 | 7 | 3864 |
| 32 | 898 | 6 | 4533 |
| 31 | 855 | 5 | 4779 |
| 30 | 900 | 4 | 5758 |
| 29 | 1033 | 3 | 6094 |
| 28 | 945 | 2 | 6164 |
| 27 | 1067 | 1 | 6072 |
| 26 | 1175 | | |
| 25 | 1184 | | |
| 24 | 1244 | | |

Table S3. Size of k-shells for asymmetrical and symmetrical networks (continued)

| asymmetrical network | | symmetrical network | |
| --- | --- | --- | --- |
| $k_s$ | N | $k_s$ | N |
| 23 | 1279 | | |
| 22 | 1330 | | |
| 21 | 1371 | | |
| 20 | 1474 | | |
| 19 | 1560 | | |
| 18 | 1703 | | |
| 17 | 1665 | | |
| 16 | 1755 | | |
| 15 | 1875 | | |
| 14 | 1878 | | |
| 13 | 2040 | | |
| 12 | 1997 | | |
| 11 | 2190 | | |
| 10 | 2246 | | |
| 9 | 2237 | | |
| 8 | 2240 | | |
| 7 | 2317 | | |
| 6 | 2319 | | |
| 5 | 2242 | | |
| 4 | 2176 | | |
| 3 | 2046 | | |
| 2 | 1713 | | |
| 1 | 1355 | | |